\documentclass[preprint,number]{elsarticle}
\usepackage{graphicx}
\usepackage{amsmath}
\usepackage{amssymb}
\usepackage{color}
\usepackage{psfrag}

\newcommand {\new} {}

\begin{document}
\title{Reconstruction of a random phase dynamics network from observations}
\author[up,unn]{A. Pikovsky}
\ead{pikovsky@uni-potsdam.de}
\address[up]{Institute for Physics and Astronomy, 
University of Potsdam, Karl-Liebknecht-Str. 24/25, 14476 Potsdam-Golm, Germany}
\address[unn]{Research Institute for Supercomputing, Nizhni Novgorod State University,
Gagarin Av. 23, 606950, Nizhni Novgorod, Russia}

\begin{abstract}
We consider networks of coupled phase oscillators of different complexity: Kuramoto-Daido-type networks, generalized
Winfree networks, and hypernetworks with triple interactions. For these setups an inverse problem of reconstruction of the 
network connections and of
the coupling function from the observations of the phase dynamics is addressed. We show how a reconstruction based
on the minimization of the squared error can be implemented in all these cases. Examples include random networks with full
disorder both in the connections and in the coupling functions, as well as networks where the coupling functions
are taken from experimental data of electrochemical oscillators. The method can be directly applied to asynchronous dynamics
of units, while in the case of synchrony, additional phase resettings are necessary for reconstruction. 
\end{abstract}

\maketitle
\section{Introduction}
Understanding connectivity of a complex dynamical network is a challenging inverse problem in
nonlinear dynamics, with many interdisciplinary applications  in neurosciences, ecology, climate dynamics,
and other fields, 
see~\cite{Lehnertz-11,Siguhara_atel-12,Tsonis-Swanson-12}. Many networks, including physically relevant ones like
arrays of lasers and Josephson junctions, power grids, etc., consist of oscillating elements. In this case the essential properties
of the dynamics, especially for weak coupling, can be described in terms of the phases. From the inverse problem 
perspective, one faces a task to understand the connectivity of the network from the observations of the oscillatory dynamics of the nodes.
Quite advanced methods have been developed here for a few coupled 
oscillators~\cite{Rosenblum-Pikovsky-01,Palus-Stefanovska-03,Tokuda-Jain-Kiss-Hudson-07,Kralemann-08,Penny_et_al-09,%
Kralemann-Pikovsky-Rosenblum-11,Kralemann_etal-13,Kralemann-Pikovsky-Rosenblum-14}. For larger networks
of oscillatory units, approaches based on the information measures and Bayesian estimations have been 
developed~\cite{Lobier_etal-14,Wilmer_etal-12,PhysRevE.86.061126,Stankovski12,Ota-Aoyagi-14}. In a recent 
paper~\cite{Alderisio_etal-17}, a method allowing to extract some properties of a network but not the exact strengths of coupling,
has been suggested.

In this paper, we present a rather general 
approach to reconstruct the coupling phase dynamics functions in a 
network of oscillators. It applies also in the cases where all the couplings are different and random,
both to full and to sparse networks. We present several variants of the technique: (i) for the simplest
case of a Kuramoto-Daido
phase dynamics, where the interactions depend on the differences of the phases only; (ii) for 
generalized Winfree-type interactions
which are pairwise but otherwise general functions of the phases; and (iii) for hypernetworks with triple interactions.

\section{Hierarchy of phase dynamics models}
\label{sec:hpdm}
Phase model 
or phase reduction is a general theoretical tool to describe weakly 
coupled oscillators~\cite{Kuramoto-84,Pikovsky-Rosenblum-Kurths-01,Izhikevich-Ermentrout-08}.
It is based on the asymptotic stability of the amplitude of a self-sustained oscillator and on the neutral stability
of its phase. This allows one to assume the amplitudes of coupled oscillators to be nearly constants, and to reduce
the dynamics to that of the phases. In the most general setup, the phases $\theta_i$ of $N$ uncoupled oscillators
just rotate with their natural frequencies
\begin{equation}
\dot\theta_k=\omega_k,\qquad k=1,\ldots,N\;.
\label{eq:p1}
\end{equation}
In the case of coupling they are interdependent:
\begin{equation}
\dot\theta_k=\omega_k+F_k(\theta_1,\ldots,\theta_N)\;.
\label{eq:p2}
\end{equation}
\new{Quite generally, the coupling functions can be represented by Fourier series
\begin{equation}
F_k(\theta_1,\ldots,\theta_N)=\sum_{m_1,m_2,\ldots,m_N}C_k[m_1,m_2,\ldots,m_N]\exp[i(m_1\theta_1+\ldots+m_N\theta_N)]
\label{eq:f1}
\end{equation}
(Note that generally $C_k[0,0,\ldots,0]\neq 0$, what means that there is a contribution of coupling to the natural frequency term;
see Ref.~\cite{Kralemann-08} for a detailed discussion of this issue.)
The number of unknown parameters $C_k[m_1,m_2,\ldots,m_N]$ is very large; however, in many 
cases one expect that only a few of them are non-zero.}

Quite often one considers situations simpler than \eqref{eq:p2}, with pairwise coupling between the 
oscillators only. In this case, in the
first approximation in the coupling strength, one gets a pairwise coupling in terms of the phases 
as well~\cite{Kuramoto-84,Pikovsky-Rosenblum-Kurths-01}:
\begin{equation}
\dot \theta_k=\omega_k+\sum_{j=1,j\neq k}^N Q_{kj}(\theta_j,\theta_k),\qquad 1\leq k\leq N\;.
\label{eq:p3}
\end{equation}
\new{(in this case the general Fourier coefficients \eqref{eq:f1} can be represented as $C_k[m_1,m_2,\ldots,m_N]=\sum_{j\neq k}
\tilde{C}_{kj}(m_j,m_k)$.)}
A particular example of this class is the Winfree model, where the coupling terms $Q_{kj}$ decompose as
\begin{equation}
Q_{kj}(\theta_j,\theta_k)=S(\theta_k) g(\theta_j)\;,
\label{eq:p4}
\end{equation}
where function $S(\theta_k)$ is the phase response curve of oscillator $k$ and function
$g(\theta_j)$ is the driving force
from oscillator $j$. \new{In this case Fourier coefficients $ \tilde{C}_{kj}$ additionally factorise: $ \tilde{C}_{kj}(m_k,m_j)=\tilde{C}_{kj}^S(m_k)
\tilde{C}_{kj}^g(m_j)$.}

Recently, more general cases of phase coupling have been considered~\cite{Kralemann-Pikovsky-Rosenblum-11}. 
The next level of complexity in comparison
to the pairwise coupling networks~\eqref{eq:p3} are the hypernetworks with triple coupling 
terms~\cite{Komarov-Pikovsky-13,Komarov-Pikovsky-15b,Bick-Ashwin-Rodrigues-16}
\begin{equation}
\dot \theta_k=\omega_k+\sum_{j=1,j\neq k}^N
\sum_{l=1,l\neq k,l> j}^N
G_{kjl}(\theta_k,\theta_j,\theta_l),\qquad 1\leq k\leq N\;,
\label{eq:p5}
\end{equation}
\new{where the condition $l>j$ is needed to count each triplet $kjl$ only once. Here Fourier coefficients have the form
$$C_k[m_1,m_2,\ldots,m_N]=\sum_{j\neq k}\sum_{l=1,l\neq k,l> j}
\tilde{C}_{kjl}(m_j,m_k,m_l).$$}
Such terms also arise for an originally pairwise coupling between the oscillators, 
in higher orders of the phase reduction.

The phase model with a pairwise coupling \eqref{eq:p3} can be further reduced under an assumption that
the natural frequencies of the oscillators are large and close to each other, and the coupling is very weak. In this case
one performs an averaging over the fast period of oscillations, so that the resulting coupling terms depend
on the phase differences only. This form of the phase dynamics is often called Kuramoto-Daido model:
\begin{equation}
\dot\theta_k=\omega_k+\sum_{j=1,j\neq k}^N \Gamma_{kj}(\theta_j-\theta_k),\qquad 1\leq k\leq N\;.
\label{eq:p6}
\end{equation}
\new{This model corresponds to an additional diagonal condition on the Fourier coefficients:  $C_k[m_1,m_2,\ldots,m_N]=\sum_{j\neq k}
\tilde{C}_{kj}(m_j)\delta[m_k+m_j]$.}

In this paper we will discuss methods for reconstructing the phase dynamics models (in the order of their complexity: 
\eqref{eq:p6},\eqref{eq:p3},\eqref{eq:p5})
from the observations of the dynamics. We will assume here that the phases can be extracted from the time series of the oscillations,
as described, e.g., in Refs.~\cite{Kralemann-08,Kralemann-Pikovsky-Rosenblum-11}.

\section{Reconstruction of phase models}

\subsection{Kuramoto-Daido model}

We consider a set of $N$ phase oscillators $\theta_i(t)$ governed by Eqs.~\eqref{eq:p6}.
We assume that the coupling functions $\Gamma_{kj}$ 
are arbitrary (e.g., random), as well as the frequencies $\omega_k$,
and the overall dynamics is asynchronous (the case of synchronous dynamics will be treated below separately).
The problem is, how to reconstruct the coupling functions from the multivariate time series $\{\theta_k(t)\}$.
Below for simplicity of presentation, we describe how to reconstruct the coupling functions acting on the first
oscillator, $\Gamma_{1j}$, all other components are obtained similarly. 

As the first step we calculate from the data the time derivative $\dot\theta_1$ (the most common approach here is to use
Savitzky-Golay filtering, see \cite{Ahnert-Abel-07}).
Then, we represent the unknown functions $\Gamma$ as a Fourier series
\begin{equation}
\Gamma_{1j}(x)=\sum_{m=1}^M \left(C_{1j,m}\cos m x+S_{1j,m}\sin m x\right)\;.
\label{eq:2}
\end{equation}
\new{Here $M$ is the maximal order of the harmonics in the reconstruction. It is an important parameter of the method
to be discussed in details below.}
At each observation point $t_n$, the following equation should be valid
\begin{equation}
\dot\theta_1(n)=\omega_1+\sum_{j=2}^N \sum_{m=1}^M \left[C_{1j,m}\cos m (\theta_j(n)-\theta_1(n))
+S_{1jm}\sin m(\theta_j(n)-\theta_1(n))\right]\;.
\label{eq:3}
\end{equation}
If we have $L$ observation points, then Eq.~\eqref{eq:3} constitutes $L$ relations for $2M(N-1)+1$ 
unknown parameters $\{C_{1j,m},S_{1j,m},\omega_1\}$ in \eqref{eq:3}. Because these parameters enter linearly, 
determining them can be easily performed with a least squared error approach, i.e. we minimize the mean squared error
\begin{equation}
\sum_{n=1}^L \left(\dot\theta_1(n)- \omega_1-\sum_{j=2}^N \sum_{m=1}^M \left[C_{1j,m}\cos m(\theta_j(n)-\theta_1(n))
+S_{1jm}\sin m(\theta_j(n)-\theta_1(n))\right]\right)^2\;.
\label{eq:mse}
\end{equation}
The unknown parameters $\{C_{1j,m},S_{1j,m},\omega_1\}$ are obtained
by solving the corresponding
system of linear equations, e.g., via singular value decomposition~(SVD, \cite{Press-Flannery-Teukolsky-Vetterling-89}, Ch. 15).
The condition for a successful solution of the problem is $L\geq  2M(N-1)+1$. \new{This relation can be applied if  the data
are not oversampled, i.e. if the observations are independent enough. Lack of independence leads to an underdetermined problem, 
what  can be identified by small extremely singular values appearing in course of SVD.}

\subsubsection{Example 1.}
\new{We start with an elementary toy example, where we just construct a network according to the rule~\eqref{eq:2}. It will serve
to clarify the role of parameter $M$, the maximal harmonics order in the reconstruction, in comparison with the maximal order
of the harmonics in the equations.}

In this example we generated a fully connected random network of $N=16$ coupled phase oscillators~\eqref{eq:p4}
by sampling the natural frequencies from the Gaussian distribution  $\mathcal{N}(0,1)$. The coupling
functions have been generated by taking random Fourier harmonics up to order $P$:
\[
C_{kj,m},\;S_{kj,m}=\begin{cases}
\frac{1}{2\sqrt{10}m}\mathcal{N}(0,1), & m\leq P,\\
0& m>P.
\end{cases}
\]
Here $P=5$ is the maximal order of the Fourier components of the coupling.
The dynamics of the network was asynchronous. The phases were sampled with time step $0.01$, and
their time derivatives have been calculated from the time series
using the Savitzky-Golay filter (\cite{Press-Flannery-Teukolsky-Vetterling-89}, Ch. 14). 
The time series was stored with time step $\Delta t=1$, \new{to avoid oversampling}. 

\begin{figure}[!hbt]
\centering
\includegraphics[width=0.6\columnwidth]{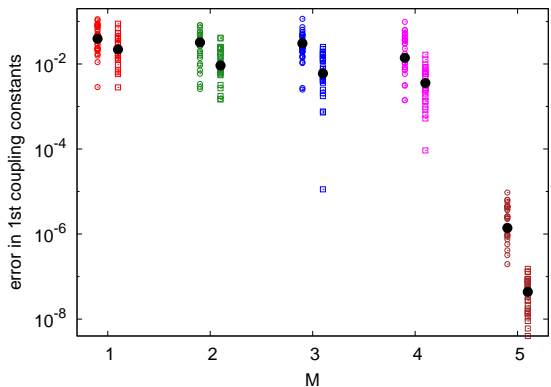}
\caption{Errors in the reconstructed coupling  constants$(C_{1j,1},\;S_{1j,1})$  as functions of the maximal Fourier harmonics $M$ 
(the plots are slightly shifted
horizontally for better visibility). Open circles: short time series $L=200$; squares: the same for $L=2000$. 
Black filled circles show median values
of the corresponding errors. }
\label{fig:reck}
\end{figure}

In the reconstruction procedure as described above, there are two essential parameters: the length of
the time series $L$ and the maximal order of the reconstructed Fourier harmonics $M$. In the case
$M\geq P$, one expects a good reconstruction, and indeed in this case we found that even for a rather short time series $L=200$, 
the maximal
error in determining the coupling constants $(C_{1j,m},\;S_{1j,m})$ was less than $10^{-5}$ (Fig.~\ref{fig:reck}). 

However, if the number of the harmonics assumed in the reconstruction was insufficient, i.e. $M<P$, then the reconstruction
was not perfect. We illustrate this in Fig.~\ref{fig:reck}. One can see that the quality of reconstruction significantly improves for
longer time series, and when more Fourier harmonics in \eqref{eq:2} are included.

\subsubsection{Example 2.}
\begin{figure}[!hbt]
\centering
\includegraphics[width=0.6\textwidth]{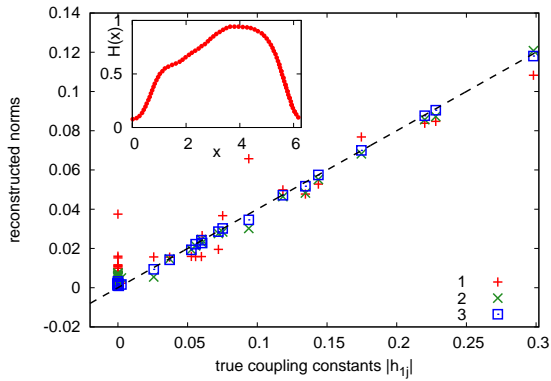}
\caption{
Reconstructed norms of coupling functions in~\eqref{eq:exp-1}, for the 1st oscillator in the network. Red pluses: 
reconstruction with $M=1$, green crosses: $M=2$, blue squares: $M=3$.
Dashed line helps to
recognize a linear relation between these quantities; \new{ the slope of this line allows estimating $\lVert H\rVert$}. Inset: Experimentally obtained coupling function $H(x)$ 
for relaxation electro-chemical oscillators, 
after~Refs.~\cite{Kiss-Zhai-Hudson-05,Kiss-Zhai-Hudson-06}.}
\label{fig:ex_cf2}
\end{figure}

In a realistic situation, the coupling functions $\Gamma(\cdot)$ are represented by an infinite Fourier series, where
however higher harmonics are quite small. In this situation, choosing a moderate value
for the maximal reconstruction mode number $M$
yields a good result. We illustrate this by employing the coupling function from the 
experiments~\cite{Kiss-Zhai-Hudson-05,Kiss-Zhai-Hudson-06}, where it has been obtained by fitting the observed dynamics
of two electrochemical oscillators. We take the coupling function $H$ corresponding to the relaxation chemical oscillations
(case (c) of Fig. 2 in \cite{Kiss-Zhai-Hudson-05}, see inset in Fig.~\ref{fig:ex_cf2}), and constructed a random network of Kuramoto-Daido type
of $N=32$ phase oscillators, coupled via $H$. The randomness in the networks is not related to the form of the coupling function, but
to the coupling amplitudes $h_{kj}$ and random phase shifts $\Phi_{kj}$: 
\begin{equation}
\dot\theta_k=\omega_k+\sum_{j=1,j\neq k}^N h_{kj} H(\theta_j-\theta_k-\Phi_{kj}),\qquad 1\leq k\leq N\;.
\label{eq:exp-1}
\end{equation}
Here the coupling constants are sampled from a  Gaussian distribution; additionally some links are excluded (so that the network is not 
fully connected):
\begin{equation}
h_{jk}=\begin{cases} 0.125 \mathcal{N}(0,1)&\text{ with probability } 0.7,\\ 0 &\text{ else, }
\end{cases}
\end{equation}
and the phase shifts $\Phi_{jk}$ are uniformly distributed  in $(0,2\pi)$.

The number of modes $M$ used in the reconstruction is the essential parameter of this problem.
 Fig.~\ref{fig:ex_cf2} shows the reconstructed norm $\lVert \Gamma_{1j}\rVert$ 
 of the coupling function, for different values of $M$.
This norm is plotted vs the coupling constant $|h_{1j}|$, which is proportional to the norm  $|h_{1j}|\cdot\lVert H\rVert$ of the true coupling
function . Thus,  a linear relation between these quantities indicates for a good reconstruction.
One can see that starting from $M=3$, the quality of the reconstruction is very good.

\subsection{Generic pairwise coupling}
Here we consider the case of a general pairwise-coupled network of phase oscillators~\eqref{eq:p3}. The basic
idea is the same as above, the only difference is in a more general representation of the coupling functions
through a double Fourier series. Let us consider harmonics up to order $P$, and denote the $2P+1$ basis functions as
\begin{equation}
f_{2n-1}(x)=\cos n x,\quad f_{2n}(x)=\sin nx,\quad f_{2P+1}=1\;,\quad n=1,\ldots,P\;.
\label{eq:bf}
\end{equation}
Then, the entries on the r.h.s. of Eq.~\eqref{eq:p3} can be approximated by 
\begin{equation}
Q_{1j}(x,y)=\sum_{m=1}^{2P} \sum_{l=1}^{2P+1} q_{jml}^{(1)} f_m(x)f_l(y)\;.
\label{eq:w2}
\end{equation}
In this representation we assume that the forcing does not contain constant terms  (in presence of such terms, 
one cannot distinguish, which 
driving oscillator produces them), but the coupling function can contain terms 
depending on the driving phase only. Thus, each coupling function $Q$ is characterized by 
$(2P+1)2P$ real numbers. In the reconstruction we use the same representation \eqref{eq:w2}, however the maximal order
of the harmonics $M$ may deviate from $P$.

The unknown parameters $\{q_{jml}^{(1)},\tilde\omega_1\}$ appear linearly in expression~\eqref{eq:w2}.
Thus, for $L$ observations
of the phases at time instants $t_n$, minimization of the squared error
\begin{equation}
\sum_n\left(\dot\theta_1(n)-\sum_{j=2}^N \sum_{m=1}^{2M} \sum_{l=1}^{2M+1} q_{jml}^{(1)} f_m(\theta_j(n))f_l(\theta_1(n)) -
\tilde\omega_1\right)^2
\label{eq:w4}
\end{equation}
leads to a linear problem of finding the unknown parameters. The length of the time series 
should be at least $(N-1)2M(2M+1)+1$ to ensure 
solvability, \new{provided that there is no oversampling}. 

\begin{figure}[!thb]
\centering
\includegraphics[width=0.6\columnwidth]{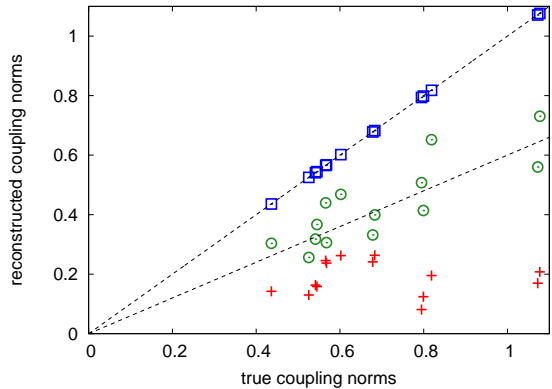}
\caption{Results of the reconstruction of the coupling constants for the 1st oscillator in the random network~\eqref{eq:p3},
for $L=5000$: Norms of the reconstructed couplings $|Q_{1j}|^2$  vs the original ones. Pluses: $M=1$, circles: $M=2$, squares: $M=P=3$. 
Dashed lines help to see validity of a linear relation between the reconstructed and true norms.}
\label{fig:w-er0}
\end{figure}

In the illustration of this approach we consider a fully connected random network with $N=16$. The frequencies are
chosen from the Gaussian distribution $\mathcal{N}(10,1)$ (contrary to the Kuramoto-Daido-type model, here the natural frequencies cannot 
be shifted arbitrary, thus one has to assume them to be relatively large compared to the coupling terms).
The coupling constants are chosen as Fourier series with $P=3$ harmonics, the latter are sampled from a Gaussian distribution. The results
of the reconstruction for $M=1,2,3$ are shown in Fig.~\ref{fig:w-er0}. One can see that for $M=P=3$ the reconstruction
is pretty good, while for an approximation with an insufficient number of 
harmonics, errors are significant. Here potentially false positives and false negatives may appear from the reconstruction 
(cf.~\cite{Alderisio_etal-17}), these features are under ongoing study.

\subsection{Reconstruction of a hypernetwork with triple couplings}
\label{sec:h}
Here we briefly discuss reconstruction of a hypernetwork~\eqref{eq:p5}. In terms of the basis functions~\eqref{eq:bf}, one can represent
the r.h.s. of \eqref{eq:p5}, similarly to \eqref{eq:w2}, as
\begin{equation}
G_{1jl}(x,y,z)=\sum_{m=1}^{2P+1} \sum_{p=1}^{2P} \sum_{s=1}^{2P}C_{jlmps}^{(1)} 
f_m(x)f_p(y)f_s(z)\;.
\label{eq:h2}
\end{equation}
The squared error can be written, similarly to~\eqref{eq:w4}, as
\begin{equation}
\sum_n\left(\dot\theta_1(n)-\tilde\omega_1-\sum_{j=2}^N \sum_{l=j+1}^N
\sum_{m=1}^{2M+1} \sum_{p=1}^{2M} \sum_{s=1}^{2M}
C_{jlmps}^{(1)} f_m(\theta_1(n))f_p(\theta_j(n))f_s(\theta_l(n))\right)^2\;.
\label{eq:h3}
\end{equation}
As a result, the unknown parameters $\{C_{jlmps}^{(k)},\tilde\omega_k\}$ can be found by singular value decomposition.
The number of these parameters for each $k$ is $(2M)^2(2M+1)(N-1)(N-2)+1$, this number defines the minimal length of the time series
required for the reconstruction. In a numerical example with $N=12$ and $P=M=2$, we generated a random hypernetwork similarly 
to examples above. This network could be reconstructed with absolute 
accuracy $< 5\cdot 10^{-3}$ with the method described, for $L=5000$ observation points.

\section{Synchronous dynamics}
\label{sec:sd}
In many cases coupled phase oscillators synchronize. In this regime, the mean frequencies coincide and the phase differences
are nearly constant (or exactly constant for the Kuramoto-Daido-type model~\eqref{eq:p6}). In this case a long observation
of the dynamics does not provide any information about the coupling terms. 
\new{This can be reformulated also as follows: a good reconstruction of the dynamics on the high-dimensional
torus, spanned by the phases, requires a good ``filling'' of this torus with trajectories. This is possible in the asynchronous case only;
for synchrony just
one periodic trajectory does not ``fill'' the torus and the reconstruction fails.}
Instead, one can effectively ``fill'' the phase space, and to get an information about the coupling terms,
via observations of the transients to the synchronous state. For a large network, many observations of the transients are necessary,
according to the estimations of the number of independent measurements presented above. These transients can be
implemented, e.g., via random resettings of the phases, as suggested in Ref.~\cite{Levnajic-Pikovsky-11}. 

\begin{figure}[!thb]
\centering
\includegraphics[width=0.46\textwidth]{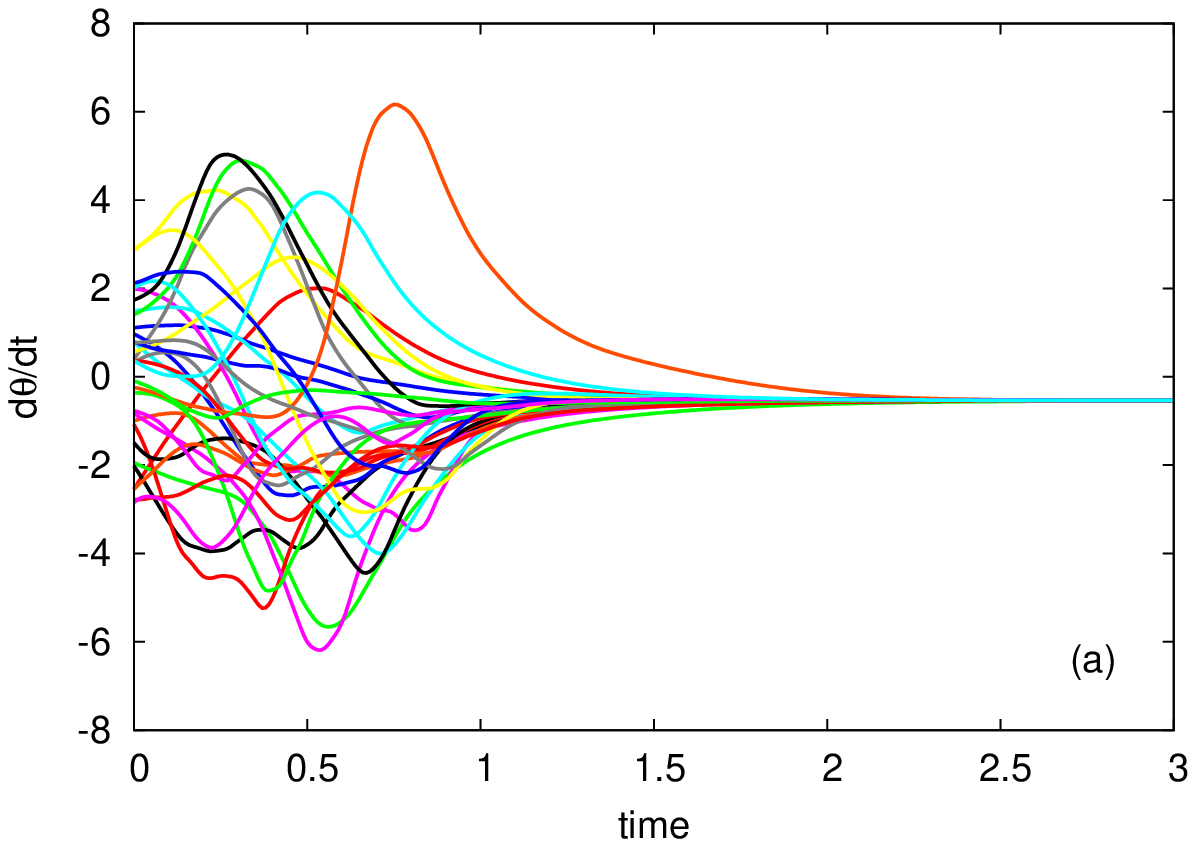}\hfill
\includegraphics[width=0.46\textwidth]{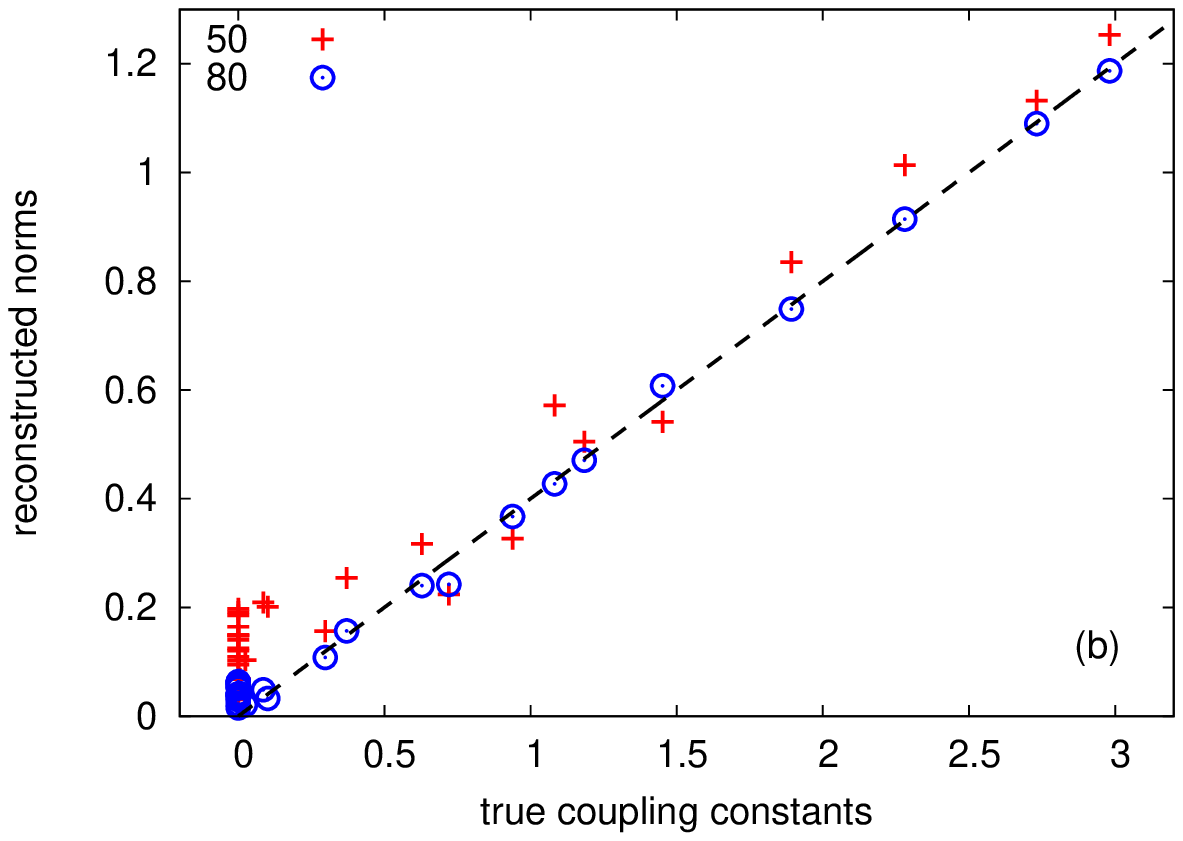}
\caption{Panel (a): Transient to a synchronized state where all the phases are locked \new{and instantaneous frequencies
$d\theta_i/dt$ coincide}.
Panel (b): Reconstructed norms of the coupling terms (with the maximal harmonics number $M=3$), vs norms of the true
coupling constants $|h_{kj}|$ in \eqref{eq:exp-1}. Pluses: 50 resettings; circles: 80 resettings. Dashed line helps to
recognize a linear relation between these quantities.}
\label{fig:ex_syn}
\end{figure}

We illustrate this approach with  a random network of Kuramoto-Daido type, where coupling functions are taken from 
experiments~\cite{Kiss-Zhai-Hudson-05,Kiss-Zhai-Hudson-06} (cf. Fig.~\ref{fig:ex_cf2} above). In a network of $N=32$
elements,
constructed like in the example of Fig.~\ref{fig:ex_cf2}, but where
all the coupling amplitudes $h_{ij}$ have the same sign, and all phase shifts $\Phi_{jk}$ are the same, 
one easily achieves synchronization (Fig.~\ref{fig:ex_syn}(a)).
Just an observation of one process like in Fig.~\ref{fig:ex_syn}(a), is insufficient for the reconstruction. However,   many
such transients, obtained via random resettings of the phases from the synchronized state, yield enough independent data points. 
The quality of reconstruction crucially depends on the number of resettings, as illustrated in Fig.~\ref{fig:ex_syn}(b). While for 50 resettings
the reconstruction is relatively poor (the values significantly deviate from a linear dependence), with many false 
positives (significantly non-zero values for the coupling norms obtained for non-existing links),
already for 80 resettings the quality is rather good (the relation between the norms is linear with high accuracy).

\begin{figure}[!thb]
\centering
\includegraphics[width=0.7\textwidth]{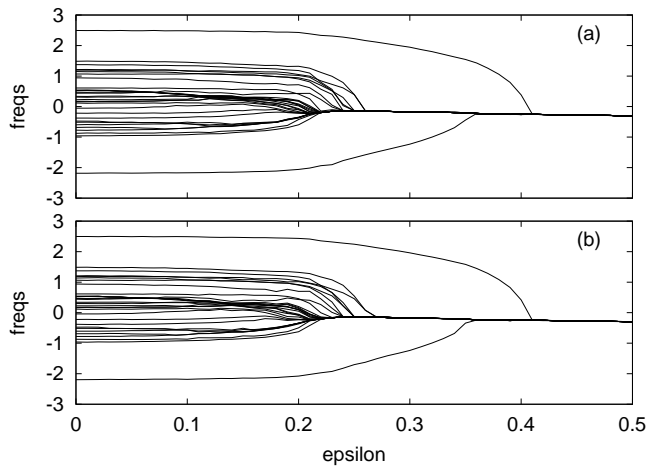}
\caption{Bifurcation diagrams for the network of oscillators, reconstruction of which is illustrated in Fig.~\ref{fig:ex_syn}.
Panel (a): reconstructed network, panel (b): original network.}
\label{fig:bif}
\end{figure}

\new{With this example we would like to discuss predictive features of the reconstruction method.  In Fig.~\ref{fig:bif} we
demonstrate, how good is the reconstruction of the bifurcation diagram of the network. We used the reconstruction 
of the synchronous network illustrated in Fig.~\ref{fig:ex_syn}. Then, we took the reconstructed equations of type
~\eqref{eq:p6}, and multiplied
all the terms, except for the constant terms $\omega_k$, with a parameter $\varepsilon$.
Then, varying $\varepsilon$, we simulated the reconstructed network equations, calculated the observed frequencies 
$\langle\dot\theta\rangle$
of the oscillators, and plotted them vs. bifurcation parameter $\varepsilon$. The same has 
been done for the original network (Fig.~\ref{fig:bif}(b)). One 
can see a good qualitative agreement between the original and the reconstructed bifurcation diagrams.}

\section{Conclusions}
In conclusion, we presented a technique to reconstruct a network from the observations of the phase dynamics of the nodes.
We illustrated it with dynamical equations of different complexity, from the simplest Kuramoto-Daido-type
dynamics, depending on the phase differences only, up to generic nonlinear triple interactions, forming a hypernetwork. 
The method works well
if the dynamics on the network is irregular. In the case of synchronous dynamics, in order to obtain 
enough independent states, one can
perform several random phase resettings and analyze the corresponding transients. 

\new{In this study, we focused on the final stage of the network dynamics analysis problem: on the reconstruction of
interactions, provided the time series of the phases are available. Certainly, the necessary preliminary steps like collecting and
preprocessing original oscillating signals, and extraction of the phases from them, may contain many potential sources 
of errors. Many aspects of extraction of the phases from the data have been discussed, also in the context of particular
experimental data, in Refs.~\cite{Tokuda-Jain-Kiss-Hudson-07,Kralemann-08,Stankovski12,Ota-Aoyagi-14}. It remains a challenging task
to combine all the steps, from the data collection to the final network reconstruction, and to analyze relevance of
potential uncertainties at each stage. The main goal of the present study was to demonstrate feasibility of the phase dynamics
reconstruction for a large variety of models, from simple Kuramoto-type ones to complex hypernetworks. Indeed, in particular applications
the complexity of the phase equations is not known in advance. Applying different variants of the reconstruction methods
described in this study, one can select the most adequate level of phase dynamics complexity, describing real interactions.
}

\new{As has been mentioned above, applicability of the approach heavily depends on the synchronization level in the network.
If synchronization is weak (what happens if the variance of the natural frequencies in the network is rather large compared
to the coupling level), the phases are to large extent independent, they cover a large part of the phase space, and the
reconstruction works well. In the opposite case of complete synchrony (what means that the variance of the natural frequencies is
rather small compared to the coupling level), the phases are highly correlated and in fact one has just one periodic orbit, 
not suitable for reconstruction. Here, as shown in Section~\ref{sec:sd}, observing transient dynamics can help. However, for this
one needs a possibility to perturb the system, moreover many different perturbations are necessary. In the intermediate case (when
the spread of the natural frequencies is of the same order as the coupling constants), one typically observes a mixed situation:
some oscillators synchronize forming clusters, other remain asynchronous. Here one can consider two situations. 
\begin{enumerate}
\item External perturbations are not possible. Here direct application of the reconstruction method described can potentially lead to
a  problem where ambiguous combination of parameters appear. In terms of the Singular Value Decomposition
approach such a situation is related to extremely small singular values. 
In this case, one could combine synchronous oscillators, forming a cluster, to 
one ``effective'' oscillatory element, and characterized it with just one phase. Then, one will have  a smaller effective network, 
where, however, all elements will be not synchronous. Here one can apply the method of reconstruction described above. The method
will, however, provide an ``effective'' coupling to a cluster, and will not allow resolving the couplings to elements of the cluster.
\item If external perturbations are possible, these can be used to destroy clusters -- one does not need to apply them to all the oscillators,
just the existing partial synchrony should be destroyed. Then, application of the approach based on the 
transients (Section~\ref{sec:sd}), will allow 
reconstructing all the connections.
\end{enumerate}
}

The method is based on the simple linear system of equations, resulting from  the minimization of squared error. 
 In this sense it is similar to that
 of Ref.~\cite{Shandilya-Timme-11}, where a general setup of high-dimensional
 dynamical systems coupled via a network of nonlinear interactions was considered. In Ref.~\cite{Shandilya-Timme-11}
 it was, however, assumed that 
 all nonlinear functions defining the local dynamics and the coupling are known, so the only unknown
 parameters, the coupling constants, could be reconstructed via minimization
 of the squared error, provided the full high-dimensional time series
 from all sites are available. In our case, no preliminary knowledge about the coupling functions is needed. 
 Moreover, we assumed all the couplings to be generally different. If one knows that the coupling functions
 in a network are the same, but still unknown, the number of unknown parameters in the reconstruction decreases. Here
an iterative method for solution of nonlinear equations resulting from the minimizing
of the squared error, suggested in Ref.~\cite{Cestnik-Rosenblum-17}, could be applied.
 Another type of error minimization, mostly suitable
 to reconstruction of very sparse networks, is known as compressive sensing, see
 Ref.~\cite{Wang-Lai-Grebogi-16} where its applications  to network 
 reconstruction are reviewed. 
 
\new{Throughout the paper we assumed that the object under study is a network of weakly coupled periodic oscillators. Here
the application of phase reduction is well justified. One can potentially avoid phase reduction by using the full reconstruction
of the equations. For lump oscillators, an efficient method for the reconstruction of the nonlinear dynamical
equations based on the ``Sparse Identification of Nonlinear Dynamics'' has been recently 
suggested in Ref.~\cite{Brunton_etal-16}. Here one heavily uses the fact, that only several nonlinear terms are important for the dynamics,
so that the equations are sparse in the space of possible functions. One can potentially extend this approach to networks, where
connections are sparse and are described by simple linear or nonlinear terms. The method, however, requires data of all the dynamical
variables from all the nodes.}

\section{Acknowledgments}
Study presented in Sects. \ref{sec:h},\ref{sec:sd} 
was supported by Russian Science Foundation [Contract No. 17 12 01534]. 

\def\cprime{$'$}

\end{document}